# AN AD-HOC SOCIAL NETWORK GENERATION APPROACH

Maurice Tchoupé Tchendji[1,2], Martin Xavier Tchembé[2] and
Armelle Linda Maténé Kakeu[2]

[1]LIRIMA - FUCHSIA team
[2]Department of Mathematics and Computer Science,
University of Dschang, Dschang, Cameroon

## ABSTRACT

*The use of social networks is still confined to infrastructure-based networks such as the Internet. However, many situations (conferences, fairs, etc.) may require the implementation and rapid deployment of an ad-hoc application for disseminating information: we call this type of application, Ad-hoc Social Network. These applications are necessarily distributed, deployable on mobile units, etc. They therefore inevitably share the same characteristics as those inherent in ad-hoc mobile networks and make them good candidates for their deployment. In this paper, by using techniques from the field of generative programming, we propose an approach to produce environments for generating such applications from their specifications in a domain-specific language. By applying this approach, we have developed SMGenerator, an environment for generating mobile ad-hoc social network applications for Android devices. Moreover by using SMGenerator, we easily generated the ConfInfo application: an ad-hoc social network application for disseminating information to participants in a scientific manifestation.*

## 1. INTRODUCTION

With the increase in recent years in the type and number of interconnected mobile electronic devices, interest in the development of mobile applications has steadily increased. The applications developed in this context are intended for various fields like health, education, teaching, etc. and used for various needs (playful, didactic, communication, etc.). An important category of these applications is used by individuals and organizations constituting a group with common interests to maintain a link among its members through the exchange and sharing of information. That is what is generally called social media or social networks for abuse. Social media is a *group of online applications that are based on the ideology and technology of Web 2.0 and it allow the creation and exchange of content generated by users* [1].

Social networks are generally designed to work in an infrastructure-based network like the Internet. However, one can imagine many situations in which one may need to create and deploy a social network-type application to meet a specific need. Such a need may arise for example if the infrastructure is non-existent or broken down, or if it is expensive (just consider the case of technologically underserved countries), etc. Such applications are generally referred to as Mobile

Ad-hoc Social Networks (MASNs) and have been the focus of several research projects, such as [2-7]. In this paper, we are interested in ad-hoc social networks based on a publish/subscribe communication model [8] in which the information exchanged in the network is grouped by subjects. We will call them SocialMANET afterwards. One can imagine for example, the need to have a dedicated tool for disseminating targeted information to the members of a conference by using exclusively the possibilities offered by their communication devices (smartphones, tablets, connected watches, etc.). Each participant will install an (*ad-hoc*) application and *subscribe* to one or more topics in line with his or her interests (relating to the conference); he or she can then *publish* and/or receive other participants' publications on this topic (see sec. 4.2 for a further presentation of this example).

Implementing from scratch an application of the type described briefly in the previous paragraph can be long and tedious, and only people with advanced programming knowledge can venture into it. If we however look closely at the structures of the applications that can be created in an ad-hoc way for a targeted dissemination of information between people who share common interests and who may very often meet in small groups (case of participants in a conference, a fair, sports events such as university games, etc.), we can see that they constitute a family of applications that can be represented by the generic software architecture (in 4 layers) presented on figure 1. Indeed, they use similar or even identical dissemination protocols, have graphical interfaces offering the same interaction styles and structured identically, etc. They only differ in terms of the topics of discussions depending on the area concerned (their titles and structure), the images/icons, and some labels used specifically on the interface of each application.

If we instantiate the architecture of figure 1 for the implementation of two (or more) ad-hoc social network applications, we can see (see sec. 2.3) that the set of layers can be partitioned into two subsets: the one made up of the so-called *common layers* (Service, Application and Communication layers) whose implementation is reused for all SocialMANETs and the one made up of the so-called *specific layers* (HMI "Human-Machine Interface" layer + part of the storage) whose implementation differs from one SocialMANET to another. The aim of this paper which is the extended version of the one titled "Une approche de génération de réseaux sociaux ad-hoc" presented at CRI'19 [9], is to propose and experiment an approach that allows the generation of SocialMANETs. The *specific layers* are specially used in the methodology adopted for this purpose; it is declined in three steps: 1) description of a generic approach for producing SocialMANET generation environments; 2) instantiation of the proposed approach for the development of *SMGenerator* which is a SocialMANET generation environment for Android platforms; 3) Experimenting with SMGenerator for the generation of *ConfInfo* (a SocialMANET for disseminating information to scientific conference participants).

At the heart of our approach is the intensive use of techniques and tools from the field of generative programming, allowing us not only to generate code from grammatical specifications, but also to integrate them to obtain a reliable and efficient software product.

**Organization of the manuscript:** section 2 presents as preliminaries some generalities about the SocialMANETs and a brief description of the dissemination protocol we used. Section 3 on the other hand presents a reminder of some code generation techniques. We discuss about the proposed approach in section 4; its instantiation for the production of the SMGenerator environment as well as an experimentation of it for the generation of the SocialMANET ConfInfo constitute the essence of section 5. Section 6 is devoted to the conclusion.

## 2. PRELIMINARIES

### 2.1. Overview of SocialMANETs: definition, operating principle and examples

We call SocialMANET, an ad-hoc and distributed application, capable to run in an ad-hoc mobile network and allowing the dissemination of information among people interested in events related to one or more topics. These topics may concern various fields (sport, study, conferences, etc.) and can be broken down if necessary into sub-topics forming a hierarchy that can be represented in tree form (see fig 3).

To receive publications on a topic, the SocialMANET user must first subscribe to it: the communication model used here is that of publish/subscribe and precisely its subject-based variant, using the dissemination protocol defined in [10]. Publish/subscribe offers a high level of decoupling in time and space between the interacting parties. The actors in a communication system based on this communication model are generally the publishers (producers of information) and the subscribers (consumers of information). Publishers populate the system with information and associate it with subjects through a *publishing action*, and the system takes care of channelling this information to those who have previously declared their interests concerning these subjects (the subscribers) through a *subscription action*. Note that a subscription to a topic belonging to a hierarchy implicitly entails a subscription to all the (sub) topics below it (*"child topics"*) in the hierarchy. This allows not only to make refined subscriptions but also to reduce the number of subscriptions needed to receive events on several topics.

In SocialMANETs, information is only disseminated during fugitive contacts among nodes (which are mobile stations and communicate by radio waves that they transmit and receive). However, the mobility of the nodes, coupled with the existence of altruistic nodes (such nodes carry publications even on topics to which they do not subscribe), allows for wide-ranging dissemination. Indeed, due to the mobility of the nodes, an information will end up with a fairly high probability of being disseminated even to those subscribers who are far from the initial place of its publication [10].

Situations that require the use of SocialMANETs may be recognized according to the following non-exhaustive list of criteria: (1) need to disseminate information in connection with a temporary circumstance; (2) grouping of individuals who can move and communicate in a restricted geographical area; (3) network infrastructure for communication that is costly or out of order, or simply non-existent. There are many examples of such situations, some of which are listed below:

#### 2.1.1. Conferences (scientific event)

Conferences are generally structured in workshops, plenary sessions and small committee meetings (meetings of the organizing committee). Participants can be grouped into several categories, including speakers, committees, chairs, guests. With the help of a SocialMANET, the different participants can exchange information on each of the categories listed above at no cost and without effort.

#### 2.1.2. Dissemination of academic information within a faculty

Students within a faculty generally want to be informed about activities that take place on campus, including courses, tutorials, assignments, exams and the publication of results. Information about these activities is said to be academic. To allow the exchange of this information at no cost and without effort between students, the use of a SocialMANET may once again be recommended. Indeed, by moving around, a student can convey information from one point to another on campus. Moreover, since these students very often group together according to whether they are taking the same courses, participating in the same exams, etc., the power of the publish/subscribe communication model used by SocialMANETs will make exchanges systematic, allowing students to avoid surprises due to lack of information.

### 2.1.3. University Games

The University Games is a sporting event that brings together the universities and colleges of a country to compete in various disciplines on the campus of a host university. For several days, this large community of academics engages in several activities involving crowds of players, spectators and many others. Under these conditions, we can envisage the creation of a SocialMANET to allow these people to inform themselves, free of charge and without effort, of the dates and places where the various activities take place.

## 2.2. Brief description of the dissemination protocol used

The dissemination protocol used in this paper is a distributed publish/subscribe protocol executed collaboratively by the stations of an ad-hoc mobile network (all stations execute the same protocol); it has been proposed by Maurice Tchoupé et al. [10]. It is subdivided into three sub-protocols which are: (1) the subscription sub-protocol, (2) the publication sub-protocol, and (3) the dissemination sub-protocol. The first two suggest what they consist of by their rather indicative names and are relatively simple. The third constitutes the core of this protocol and makes it possible to transfer publications from one station in the network to other stations based on their centres of interest and/or their altruistic nature. It consists of two phases executed by all the stations of the network independently of each other: (i) the *detection of requirements* phase in which a station, by signalling itself in a neighbourhood, enables other stations in that neighbourhood to detect which of its requirements can be met, and (ii) the *transfer of publications* phase in which stations broadcast the publications they possess in accordance with the detected requirements of stations in their vicinity. These requirements are based on common interests with neighbouring stations and represent local publications related to those interests that they do not possess.

In practice, this protocol is executed on a virtual network layer within an existing physical network (the ad-hoc mobile network). Any station in the network participating in the SocialMANET protocol execution is autonomous and can communicate with any other station in its neighbourhood (a station's neighbourhood consists of all the stations that are within its radio range). Each station can be both publisher and subscriber. All stations ensure the dissemination of the publications they own to interested subscribers.

## 2.3. Generic architecture of the SocialMANETs

The generic architecture of a SocialMANET node (see fig. 1) consists of four layers: the Communication, Application, Service and HMI layers.

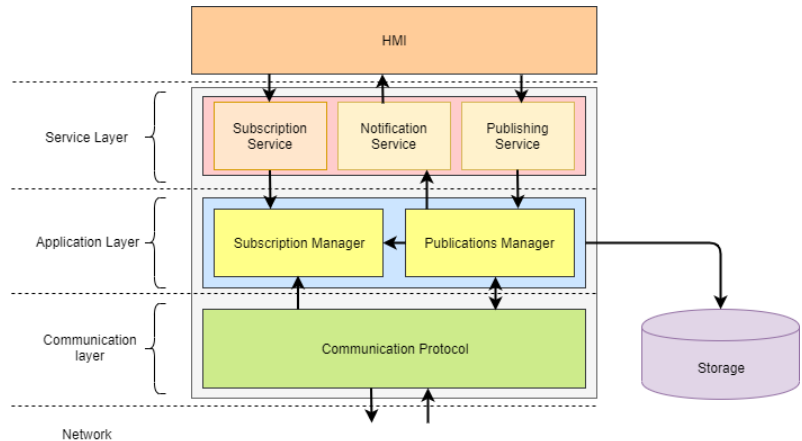

Figure 1: Generic (layered) architecture of SocialMANETs

The Communication layer ensures communications between the current node and the other nodes of the network; it is at this level that the dissemination protocol used [10] is implemented. On reception of a publication likely to interest the current node, it transfers it to the "Publication manager", located in the Application layer. The latter implements the business logic of the application through two components whose very suggestive names suggest their roles: the *Manager of publications* and the *Manager of subscriptions*. The Service layer, on the other hand exposes the Subscription and Publication services that can be invoked by the user through the HMI, in particular to subscribe to a subject (Subscription Service) or to publish information (Publication Service); it also exposes the Notification Service which can be invoked by the system to notify the user of the reception of a new publication. The HMI layer provides an interaction style for invoking the services offered by the Service layer. Thus, for a given SocialMANET, it must allow access to the current list of subjects (those to which one can subscribe) as well as to their hierarchy, to the subscription, publication and publications management (reading, archiving, deleting, etc.) functionalities.

On reading the description given above of the generic architecture of a SocialMANET, it is easy to realize that from one specific SocialMANET to another, only the implementation of the HMI layer and (to a lesser extent) the structuring of storage (the structuring of storage into directories and sub-directories reflects the hierarchy of subjects) differs. Even more, if the GUI is graphic and offers a tactile interaction style, the screens of different SocialMANETs will have approximatively the same design (same organization of the graphic components) and the only real differences will relate to the nomenclature of the subjects and their hierarchy. Figure 2 presents two HMIs of two SocialMANETs allowing the dissemination of information during a scientific event (ConfInfo) (fig. 2(a)), and the dissemination of academic information in a higher education institution (FacInfo) (fig. 2(b)). The hierarchies of the subjects which are respectively used there are represented on figure 3.

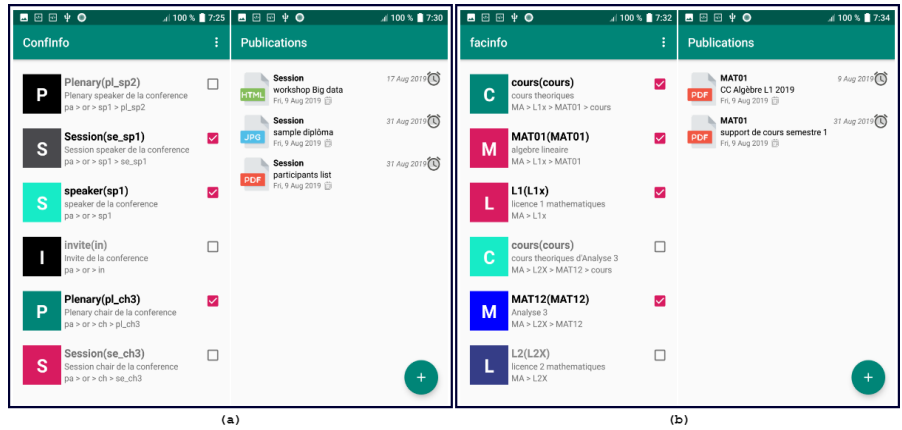

Figure 2: Overview of the GUIs of two SocialMANETs: ConInfo (a) and FacInfo (b)

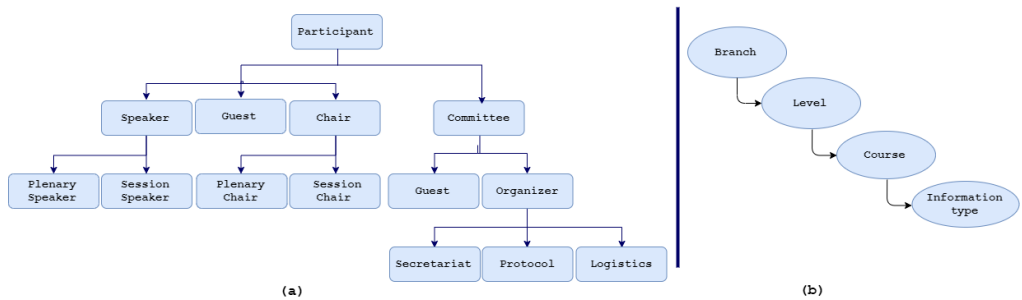

Figure 3: Subjects' hierarchies for SocialMANET ConfInfo (a) and FacInfo (b)

## 3. FOCUS ON SOME CODE GENERATION TECHNIQUES

At the end of the previous section, we noted that a specific SocialMANET differs from another only on certain aspects of the HMI relating to the subjects handled. In this section, we explore some state of the art code generation techniques.

Generative programming is a programming paradigm that aims at developing techniques to generate source code from models in order not only to reduce the development time of software products but also to guarantee their corrections with respect of the models considered [11]. The models manipulated derive from specifications which in agreement with Jean-Claude Tarby [12] and depending on the class of the problem treated, can be graphical (a UML class diagram for example), functional or based on a dedicated language or DSL (Domain Specific Language).

Carine Courb is et al [13] used the generative programming paradigm to build a generator of development environments called SmartTools. Their objectives are part of the problem of rapid and simplified design of programming languages for the exchange and processing of information. More precisely, based on a description of a language, the SmartTools platform generates a development environment containing an analyser of a concrete form of the language (parser), the associated display (pretty-printer), a syntax editor and a set of Java files facilitating the writing of semantic processing (analyses, transformations). On the other hand, SmartTools is based on object-oriented technologies: implementation in Java, use of the visitor pattern (*In object-oriented programming and software engineering, the Visitor Behavioural Design Pattern is a way to separate an algorithm from a data structure*), programming by aspects (*Aspect-Oriented Programming -AOP- is a programming paradigm that makes it possible to deal separately with cross-cutting concerns, which are often technical or business concerns that are at the heart of an application*), distribution of objects and components. The combination of these technologies has allowed the authors to propose at a lower cost, an open, interactive, uniform and scalable development platform.

For his part, OLSEN Dan R. [14], has built the SYNGRAPH system to automatically generate graphical user interfaces. This system generates interactive Pascal programs from a description of the grammar of the input language. From this grammar, it derives information about the management of physical devices. Input errors are detected and can be corrected using the automatically provided undo functions.

Because an efficient user interface requires a dialog layer that can handle multiple interaction threads simultaneously, YAP Sue-Ken et al [15] provide a notation for specifying dialogs based on context-free assigned grammars. The notation is useful both as a means of communicating the behaviour of the dialog layer to designers and as input for a compiler that outputs the interface source code.

SINGH S. et al [16] on their side propose a user interface management system (Chisel) that generates graphical user interfaces based on a detailed description of the semantic commands supported by the application. Chisel selects the interaction techniques, determines their attributes and places them on the display device screen. In doing so, it is able to take into account the properties of the device, end-user preferences and interface designer directives.

Paul Franchi-Zannettacci [17] has developed a graphical editor generator designed from a formal specification, based on a given language. For this, he describes the syntax of this language thanks to an assigned grammar and through boxes that give the graphical constraints.

Taking into account the fact that one of the declinations of our main objective is to be able to specify the subjects of a SocialMANET as well as their hierarchy, it is the specification based on DSL that seems best adapted. In this type of specification, a dedicated work environment, allowing the designer to directly manipulate the concepts of the domain is made available. This environment integrates a DSL in which the designer must write his specification. In addition, it provides access to a set of tools (interpreter, debugger, specialized editor, etc.) that allow efficient and effective use of the DSL. For example, the specialized editor must provide editing facilities (code completion, syntax highlighting, etc.) found in traditional general-purpose language editors; in addition, the embedded interpreter must implement routines to generate source code for the specification written in accordance with the DSL's syntax.

Nowadays, the design and implementation of a DSL is relatively straightforward using the Meta-DSL (Xtext [18], MPS Jetbrains [19] etc.) which are frameworks for creating DSLs; for example, the SMGenerator environment presented in section 5.1 was created with Xtext.

# 4. A GRAMMATICAL APPROACH TO GENERATE AD-HOC SOCIAL NETWORKS

## 4.1. Synoptic view of the approach

Figure 4 gives a synoptic view of the language-oriented approach we advocate for the SocialMANET generation. It can be presented as follows: (1) create a set of software components implementing the common layers of the generic architecture of the SocialMANET nodes (see fig. 1); (2) create a DSL for the specification of the concepts (topics and their hierarchy, application name, etc.) used for the implementation of the specific layers of the generic architecture of the SocialMANET nodes; (3) provide an integration tool to assemble the result of interpreting a program written following the DSL created in step 2 with the software components provided in step 1. The result is compiled into a ready-to-deploy archive.

We explain below what needs to be done in steps 2 and 3 because step 1 consists essentially in the implementation of the dissemination protocol presented in [10] and the development of the software component exposing the subscription, notification and publication services (see fig. 1). This software component is developed once and reused for the generation of different SocialMANETs. Indeed, as indicated in [20], reusability allows to considerably reduce the cost and development time of software products. Note that in practice (see sec. 4), steps 2 and 3 will be operated via an environment dedicated to the specification and generation of SocialMANETs.

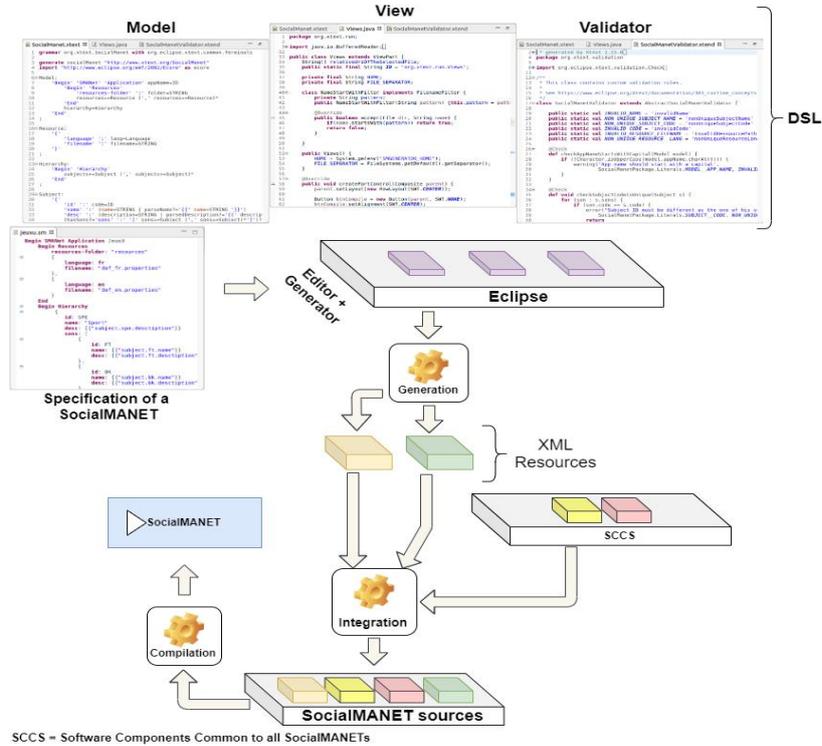

Figure 4: Synoptic view of the generation process of a SocialMANET

## 4.2. The SocialMANET DSL grammatical model and the associated interpreter

As already mentioned in section 3, the relevant concepts used for the implementation of the specific layers of a SocialMANET can be captured via a DSL of which a minimal grammar written in BNF (Backus-Naur Form) is given in listing 1. This grammar allows the designer of a SocialMANET to specify the name of the SocialMANET to be generated through the syntactic category *APP_NAME* (listing 1, line 4), the definition of the subjects and the structural relations that link them, their descriptions, through the syntactic category *SUBJECT* (listing 1, line 3).

The interpreter associated with the created DSL must output some resources (XML files) containing respectively a hierarchical description of the set of topics (see fig. 6(b)) and a set of labels to be integrated into the software components provided in step 1.

Listing 1: A Context-Free Grammar of the SocialMANET DSL

```
1 MODEL  -> APP_NAME   HIERARCHY
2 HIERARCHY -> (SUBJECT)+
3 SUBJECT -> SUBJECT_NAME SUBJECT_CODE DESCRIPTION (SUBJECT)*
4 APP_NAME -> <string>
5 SUBJECT_NAME -> <string>
6 SUBJECT_CODE -> <string>
7 DESCRIPTION -> <string>
```

## 4.3. Integration

In this step, the XML files generated in the previous step (step 2) constitute resources that must be integrated into the software component provided in step 1 (see fig. 4) to obtain the source code of the SocialMANET whose specification was made via the DSL. This code is then compiled into a ready-to-deploy archive.

# 5. APPLICATION

This section presents an example of implementation of the proposed approach for the production of SMGenerator. Also presented here is an experimentation on the use of SMGenerator to generate the Social MANET ConfInfo.

## 5.1. SMGenerator: a SocialMANET generation environment

SMGenerator is a SocialMANET generation environment and is distributed as an Eclipse (*www.eclipse.org*) plugin. The set of static software resources used by SMGenerator and representing the common layers (Service, Application and Communication layers) of the generic SocialMANET architecture (see fig. 1) is available as an Android source program at https://github.com/MartinezX21/SMGenerator. This source program lacks some resources (in the sense of Android programming) that will be provided during the integration phase. SocialMANETs generated from SMGenerator are deployable in Android devices and support several languages, thanks to an adapted grammar resulting from an extension of the minimal grammar presented in listing 1.

## 5.1.1. DSL used by SMGenerator

SMGenerator uses a DSL that we created by enriching the minimal grammar presented in listing 1 to allow it to generate multilingual SocialMANETs, i.e. supporting several languages of use. Only two languages are currently taken into account, but this simplistic model can be very easily extended to take into account all possible languages.

Listing 2 presents the Xtext grammar of the DSL used by SMGenerator. In this grammar, the syntactic category *Resource* allows to define a language of use for a new SocialMANET and the associated language file (a file containing a set of keys/values, the values being the translations into the considered language of the names and descriptions of the SocialMANET subjects). At least one resource must be added as shown in the grammar (lines 3-6). For each topic, the name and description may be static strings or keys defined in the language files and intended to be replaced by the corresponding values just before compilation, in which case the said keys must be framed by the symbols "*{{* " and "*}}*" (lines 24-27).

Listing 2: Xtext grammar of the DSL used by SMGenerator

```
1  Model:
2    'Begin' 'SMANet' 'Application' appName=ID
3      'Begin' 'Resources'
4        'resources-folder' ':' folder=STRING
5        resources+=Resource (',' resources+=Resource)*
6      'End'
7      hierarchy=Hierarchy
8    'End'
9  ;
10 Resource:
11   '{'
12     'language' ':' lang=Language
13     'filename' ':' filename=STRING
14   '}'
15 ;
16 Hierarchy:
17   'Begin' 'Hierarchy'
18     subjects+=Subject (',' subjects+=Subject)*
19   'End'
20 ;
21 Subject:
22   '{'
23     'id' ':' code=ID
24     'name' ':' (name=STRING
25             | parseName?='{{' name=STRING '}}')
26     'desc' ':' (description=STRING
27             | parsedDescription?='{{' description=STRING '}}')
28     (hasSons?='sons' ':' '['sons+=Subject(',' sons+=Subject)*']')?
29   '}'
30 ;
31 Language:
32   name=('en' | 'fr')
33 ;
```

## 5.1.2. Description of a SocialMANET on SMGenerator

After its installation, the SocialMANET generator (SMGenerator) gives access to a high level specialized editor allowing the easy specification of a new SocialMANET in an implementation of the DSL described in section 4.2, realized using the Xtext framework (*www.eclipse.org/Xtext*) which offers a solid solution for the construction of text-based DSLs under the Eclipse platform. This editor allows to write a correct specification of a new SocialMANET in a file with the extension *.sm* and to trigger the interpretation step.

To generate a SocialMANET with SMGenerator, first create a simple project in an instance of the Eclipse IDE in which the SMGenerator plugin has been installed; then create a file with the extension *.sm* in which the SocialMANET specification must be written. This file is called *SocialMANET source file* or *SM source file* for short, and the project in which it is created is called *SocialMANET project* or *SM project* for short. Figure 5(a) shows an example of a SM project with the SM source file open in the editor. The tree structure of such a project is highlighted in figure 5(b). At the root of the project, is the SM source file, a folder *resources* containing the language files for the different target languages for using the SocialMANET that you want to generate, and also a folder *src-gen* created automatically during the interpretation phase of the SM source file and containing a set of folders and files generated.

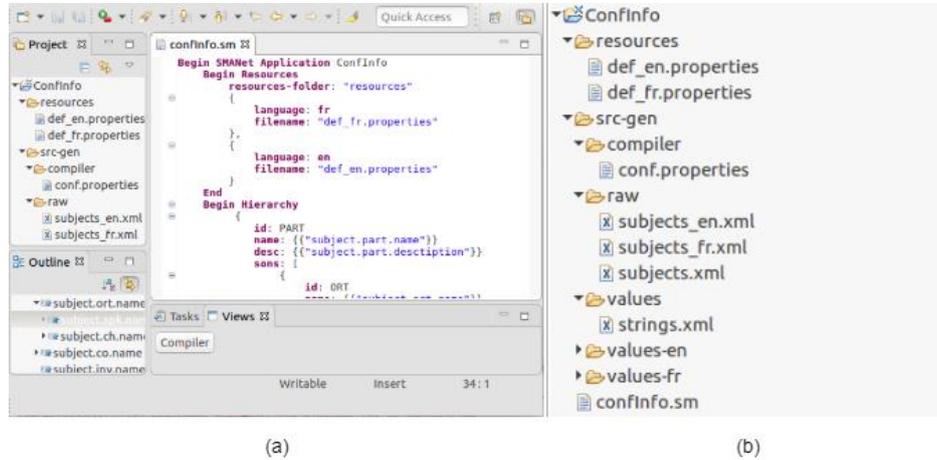

Figure 5: Description of a SocialMANET on SMGenerator

For each specified target language having the code *xx (for example, in Figure 5(b), xx corresponds to "en" for the English language and "fr" for the French language)*, a file *subject_xx.xml* is generated in the *src-gen/raw* folder and contains all the SocialMANET subjects whose names and descriptions are given in the corresponding target language. A *values-xx* folder is also created and contains the labels and static texts used by the SocialMANET and translated into the considered target language. The content of the *subject.xml* and *subject_xx.xml* files is as shown in figure 6(b). In these files, strings beginning with "*{{*" and ending with "*}}*" represent keys contained in the language files and will be replaced by the corresponding values just before compilation.

### 5.1.3. Interpretation, integration and compilation

The interpretation phase was implemented by customizing the generic interpreter generated by Xtext when creating a new DSL. More precisely, the customization consisted in writing the code to generate the XML resources to be used during the integration stage. These resources are contained in the *raw* and *values* sub-folders of the *src-gen* folder at the root of the project. In addition to these resources, a *conf.properties* file is generated in the *compiler* sub-folder of the *src-gen* folder and contains the necessary configurations for the compilation phase.

For the integration and compilation steps, we wrote a script (invoked by clicking on a button available on the GUI, see fig. 7(a)) containing a set of commands needed to compile an Android program from the generated XML resources and the static software resources. This script is responsible for merging these different resources before triggering the compilation to produce the desired SocialMANET output.

### 5.2. ConfInfo: a SocialMANET generated on SMGenerator

ConfInfo is a SocialMANET designed to facilitate the dissemination of information during a scientific event (conferences, workshops, seminars, etc.). Such an event is an important channel for information exchange between researchers. The information to be disseminated is grouped into subjects whose nomenclature reflects the categories in which the actors of the event can be classified: the speakers (plenary speaker, session speaker, and talker), the organizing committee (secretariat, protocol, and logistics), the chairs (plenary chair, session chair), the guests, etc. The hierarchy of ConfInfo topics is shown in figure 3(a). It can be seen that information published in the topic *Participant* will target all participants in the event, including speakers and members of the various committees. Similarly, information published in the topic *Organizer* will target members of the secretariat, protocol and logistics.

The generation of the SocialMANET ConfInfo was facilitated by the use of the SMGenerator environment. In summary, the use of the SMGenerator environment for ConfInfo generation (as for any other SocialMANET) is reduced to editing the subjects specification file in the proposed DSL (see figure 6(a)); an extract of one of the XML resources  generated after interpretation of this specification is presented in figure 6(b). The compilation is triggered by a click on the button "*Compiler*", as presented by figure 7(a).

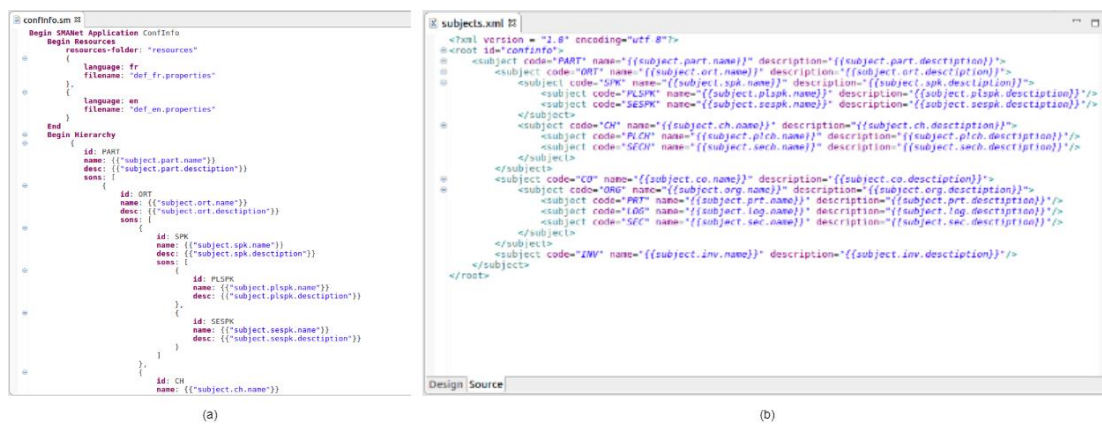

Figure 6: ConfInfo Specification File in SMGenerator (a) and correspondingly generated topic file (b)

Figure 7 presents some screen shots illustrating the generation of the ConfInfo APK. As mentioned in the previous paragraph, after editing the topic specification file, clicking on the button "*Compiler*" triggers the integration of the generated XML resources into the software component integrated into SMGenerator and then the compilation of the obtained sources (see fig. 7(a)). Once compilation is complete, the location where the APK will be copied must be specified (fig. 7(b)); finally, the generated APK is copied to the specified location, marking the end of the generation process (fig. 7(c)). Figure 2(a) shows two screen-shots of ConfInfo presenting respectively the list of topics and the publications of the topic *Session*.

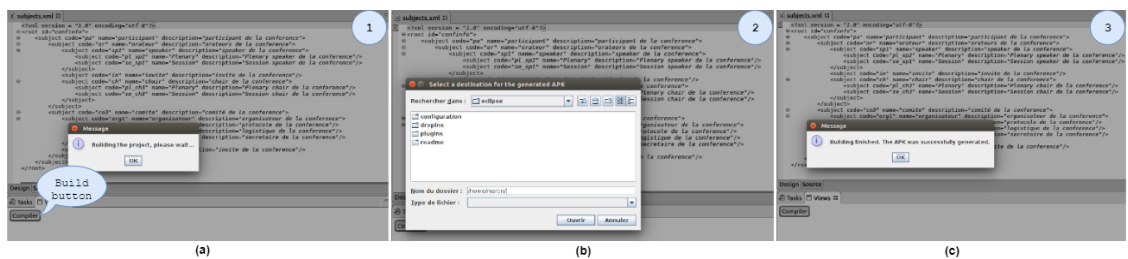

Figure 7: A summary of ConfInfo's APK generation steps

Practically, after generation of the ConfInfo APK, it is made available (*the application can be available for download on the event's website or given to participants during their registration when they arrive at the event's site*) to participants of a scientific event. They must install it in their Android devices and then subscribe to the discussion topics corresponding to their categories. Once this is done, any holder of new information to be disseminated just has to publish it in the corresponding topic and the dissemination process is immediately triggered according to the dissemination protocol described in [10]: the information is first distributed to the participants located in the vicinity of the publisher, then due to the mobility of the participants, it will finally be disseminated (with a fairly high probability) to those of them located far from the initial place of publication.

## 6. CONCLUSION

In this paper, we presented a generative approach to the production of SocialMANETs (a social network-type application that can be deployed on an ad-hoc mobile network) which in addition to drastically reduce the time devoted to their development (which is now practically limited to the time required to edit the subject specification file in the proposed DSL), allows even people without advanced programming knowledge to develop them. The implementation of the environment SMGenerator following this approach and its easy use for the generation of the SocialMANETs ConfInfo and FacInfo allowed us not only to validate the approach, but also to confirm its effectiveness and efficiency.

Note however that the proposed grammatical model (see listing 1) does not capture all the particularities that one may wish to specify for a given SocialMANET. A short-term task would be to enrich this model so that it can allow the designer of a SocialMANET to specify for example, the application's logo, colour themes, fonts to be used and their sizes, etc.

## AUTHORS

**Maurice Tchoupé Tchendji** is a Senior Lecturer in the Department of Mathematics and Computer Science at the University of Dschang (Cameroon). He received the Ph.D degree from the Universities of Yaoundé I (Cameroon) and Rennes I (France) in 2009. His research interests include process management and workflow technology, XML Information Retrieval, and Information Dissemination in Mobile Ad-Hoc Network.

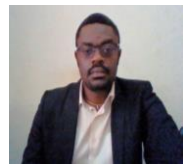

**Martin Xavier Tchembé** is a student at the University of Dschang (Cameroon). He holds a Master 2 in Network and Distributed Services obtained during the academic year 2017/2018. He is currently in his second year of thesis in the same university.

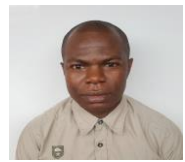

**Armelle Linda Maténé Kakeu** is a student at the University of Dschang (Cameroon). She holds a Master 2 in Network and Distributed Services obtained during the academic year 2018/2019.

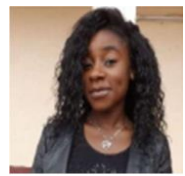